\documentclass{Interspeech}
\usepackage[none]{hyphenat}

 \interspeechcameraready

\title{Conformer-based Ultrasound-to-Speech Conversion}

\author[affiliation={1}]{Ibrahim}{Ibrahimov}
\author[affiliation={1}]{Csaba}{Zainkó}
\author[affiliation={2,3}]{Gábor}{Gosztolya}

\affiliation{Department of Telecommunications and Artificial Intelligence}{Budapest~University~of~Technology~and~Economics}{Budapest, Hungary}
\affiliation{HUN-REN--SZTE Research Group on Artificial Intelligence}{Szeged}{Hungary}
\affiliation{University of Szeged, Institute of Informatics}{Szeged}{Hungary}
\email{ibrahim@tmit.bme.hu, zainko@tmit.bme.hu, ggabor@inf.u-szeged.hu}

\usepackage{comment}

\begin{document}

\maketitle

\begin{abstract}
    


Deep neural networks have shown promising potential for ultrasound-to-speech conversion task towards Silent Speech Interfaces. In this work, we applied two Conformer-based DNN architectures (Base and one with bi-LSTM) for this task. Speaker-specific models were trained on the data of four speakers from the Ultrasuite-Tal80 dataset, while the generated mel spectrograms were synthesized to audio waveform using a HiFi-GAN vocoder. Compared to a standard 2D-CNN baseline, objective measurements (MSE and mel cepstral distortion) showed no statistically significant improvement for either model. However, a MUSHRA listening test revealed that Conformer with bi-LSTM provided better perceptual quality, while Conformer Base matched the performance of the baseline along with a 3× faster training time due to its simpler architecture. These findings suggest that Conformer-based models, especially the Conformer with bi-LSTM, offer a promising alternative to CNNs for ultrasound-to-speech conversion.

\end{abstract}

\section{Introduction}


\begin{figure*}[t]
  \centering
  \includegraphics[width=\linewidth]{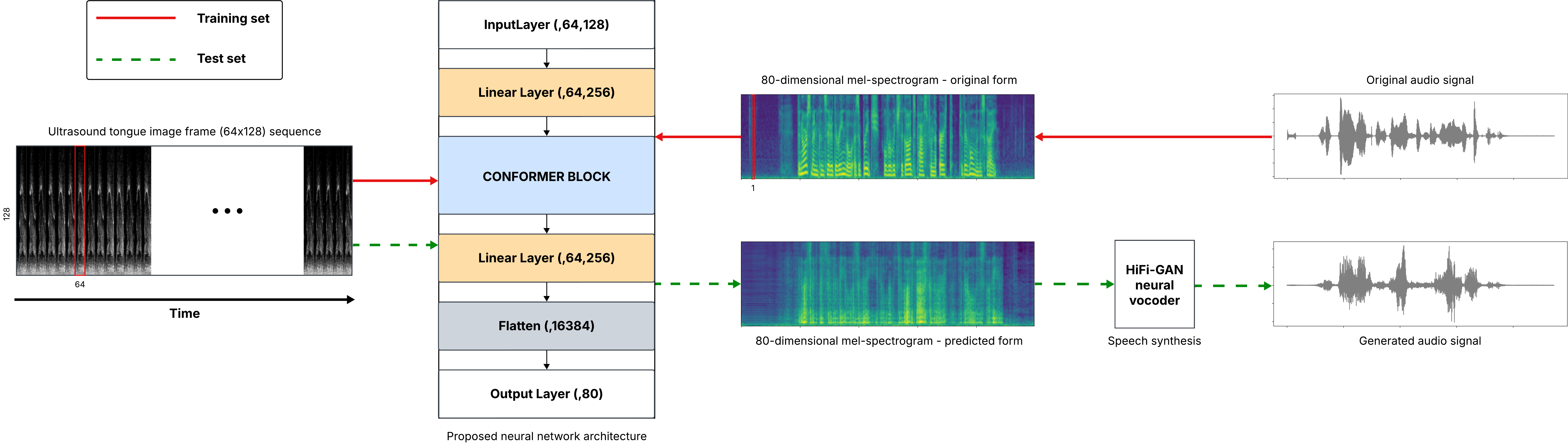}
  \caption{Schematic diagram of the ultrasound-to-speech system pipeline, showing the Conformer-based ultrasound-to-mel mapping and the HiFi-GAN vocoder for speech synthesis.  Training set with development set flows through the solid arrow, while test set flows through the dashed arrow after the training phase was done (output shape of each layer is mentioned in parentheses).}
  \label{fig:pipeline}
\end{figure*}

Speech as a communication method is a vital part of human life, involving several body parts (tongue, lips, etc.) during production. However, for various reasons (speech disorders, environment, etc.), producing audible speech is sometimes not possible or desired. To offer an alternative communication method in these scenarios, Silent Speech Interfaces (SSI) have been introduced. SSI is a technology that targets to recognize or reconstruct speech from articulatory movements. These movements can be captured using various techniques, such as surface electromyography (sEMG,~\cite{zhu2021towards,sEMG_2}), magnetic resonance imaging (MRI,~\cite{trencsenyi2024ultrasound,toutios2016articulatory}), lip videos~\cite{ribeiro21_interspeech,pandey2024melder} and ultrasound tongue imaging (UTI,~\cite{hueber2016statisticalconversion,guo2024multi}), among other modalities. Particularly, UTI is a promising technique for acquiring detailed information about tongue movement, 
while also being non-invasive and cost-effective.


Ultrasound-to-speech conversion is a task within the scope of biosignal-based speech synthesis towards real-time SSI application. To achieve this, given the established link between tongue movements and the acoustic speech signal, several methods have been developed for ultrasound-to-speech synthesis (UTS). A standard UTS workflow consists of two steps: mapping from ultrasound tongue image frame sequence (UTIF) to an intermediate representation of the speech signal (such as a mel spectrogram), and synthesizing speech from this representation using a (neural) vocoder.


Researchers have explored a wide range of DNN architectures for mapping UTIF to intermediate representations of speech. Csapó et al.~\cite{csapo17_interspeech} focused on mapping UTIF to mel-generalized cepstrum-based line spectral pair (MGC-LSP) coefficients using fully connected networks. To improve performance, subsequent research explored the use of convolutional neural networks (CNNs) combined with Long Short-Term Memory (LSTM) networks, including both single-layer recurrent and two-layer bi-directional LSTM architectures~\cite{article_csapo_lstm}. More recently, Saha et al.~\cite{saha2020ultra2speech} introduced Ultrasound2Formant Net, a 3D CNN-based architecture, to directly map UTIF to formant representations, demonstrating an alternative approach to intermediate speech representation.

Estimating an n-dimensional sound representation from UTIF in the form of mel spectrum has also been an active research direction for the ultrasound-to-speech conversion task. Kimura et al.~\cite{kimura2019sottovoce} utilized a CNN model to map 13 ultrasound images of size 128×128 to a 64-dimensional mel-scaled sound vector. They further employed a second network to refine the generated mel spectrograms, demonstrating a significant improvement in silent speech recognition accuracy from 42.5\% to 65\%. On the other hand, Csapó et al.~\cite{csapo20b_interspeech}, as part of a UTS system with a WaveGlow neural vocoder, employed a two-dimensional CNN architecture for mapping UTIF (64×128) to an 80-dimensional mel spectrum. 
Despite these advancements, there is still room for improvement in the accuracy and robustness of UTIF-to-mel mapping. 

In this work, for the first time, we implemented convolution-augmented transformer (i.e. conformer) block for the UTIF-to-mel mapping step as a part of the ultrasound-to-speech conversion task. The transformer~\cite{NIPS2017_3f5ee243}, a self-attention-based neural network, was developed to address long-sequence dependency problems. Since its introduction, the transformer architecture has inspired its application in diverse fields such as computer vision~\cite{dosovitskiy2020image}, genomics~\cite{genomics_transformer}, speech recognition~\cite{moritz2020streaming} and also SSI~\cite{SONG2023104298}. The conformer, introduced by Gulati et al.~\cite{gulati20_interspeech}, leverages the strengths of both Transformers and CNNs, combining efficient global interaction capture with effective local feature extraction. From a performance perspective the conformer model has been shown to be parameter-efficient, while effectively learning both local and global dependencies in sequential data. 

In this work we also explored a more complex architecture by adding bidirectional LSTMs (bi-LSTM) into our `Conformer base' structure. The combination of conformer blocks with bi-LSTMs has shown promising results in (biosignal-based) accelerometer-to-speech synthesis task towards SSI~\cite{KWON2024109090}. This success motivated us to also investigate its potential for UTIF-to-mel mapping. As a baseline approach, we utilized a standard 2D-CNN model.

For all UTI-to-mel mapping approaches, we generated the synthesized speech samples using the HiFi-GAN neural vocoder. 
The performance of the resulting UTS systems (see Figure ~\ref{fig:pipeline} for its pipeline) was evaluated using objective metrics, specifically Mean Squared Error (MSE) and Mel-Cepstral Distortion (MCD). We also conducted listening test using the generated audio signals for the subjective evaluation of the UTS configurations tested.

\section{Dataset}

We conducted experiments our using four participants: two females (01fi, 02fe) and two males (03mn, 04me), selected from the UltraSuite-TaL80 corpus~\cite{ribeiro2021talasynchronized}\footnote{\href{https://ultrasuite.github.io/data/tal_corpus}{https://ultrasuite.github.io/data/tal\_corpus}}. This dataset includes recording of the tongue movements using an ultrasound system called ”Micro” by Articulate Instruments Ltd. at a frame rate of 81.5 fps, simultaneously with audio recordings. The dataset contained varying numbers of recordings per speaker. We used the complete set of recordings for each speaker, specifically: 204 (01fi), 141 (02fe), 193 (03mn), and 190 (04me).

We used the scanline data of the ultrasound as the input of the neural networks, after being resized to 64×128 pixels using bicubic interpolation from its original size of 64×842, and normalizing the pixel values to the range [-1, 1]. To ensure consistent evaluation, 10 out of 24 shared read sentences (the recordings between "005\_xaud" and "014\_xaud") were reserved for testing across all speakers. The remaining dataset was partitioned into training and development sets in a 9:1 ratio per speaker.

\section{Methodology}
\begin{figure}[t]
  \centering
  \includegraphics[width=\linewidth]{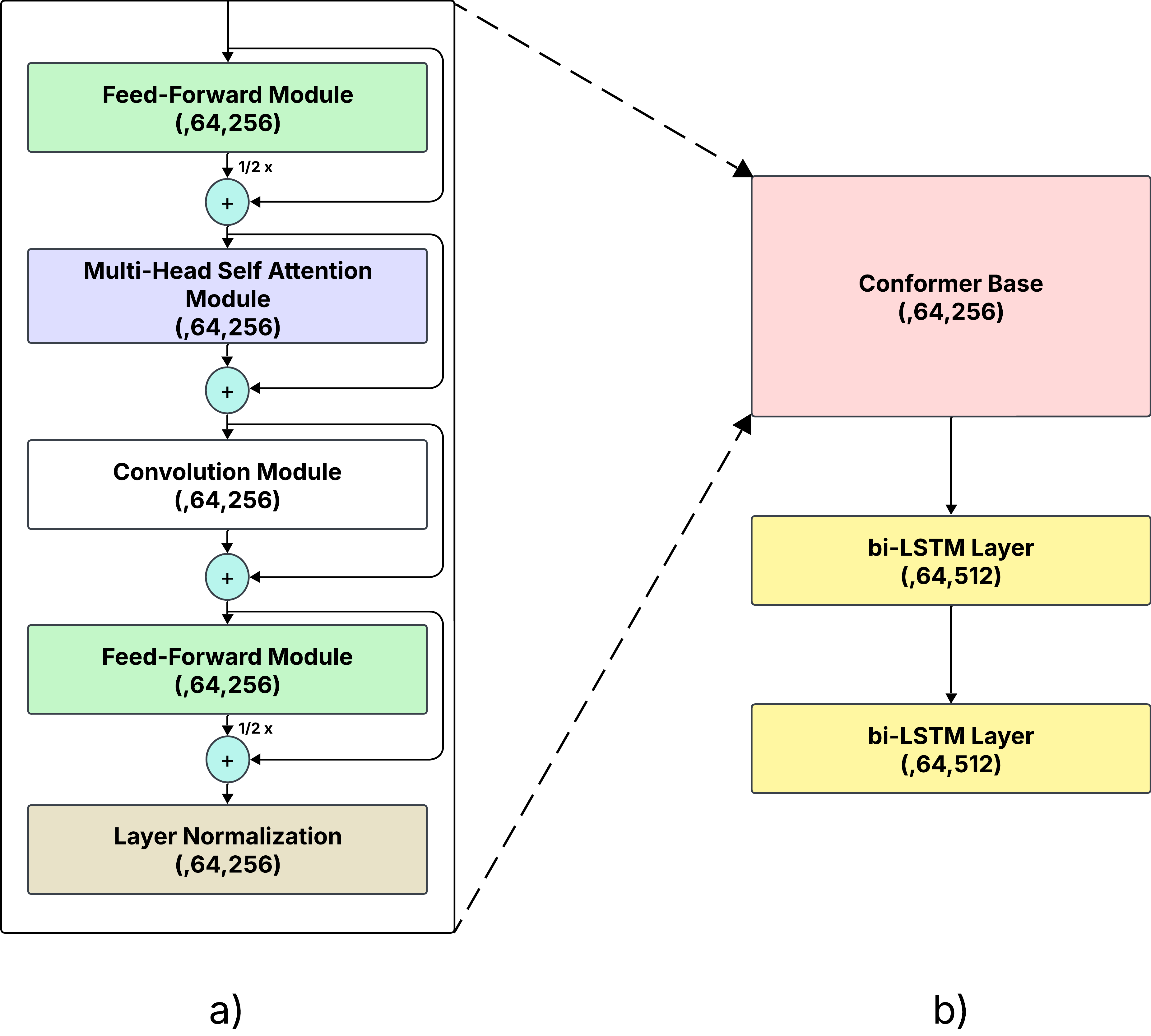}
  \caption{a) Conformer Base; b)Conformer with bi-LSTM (output shape of each layer is mentioned in parentheses).}
  \label{fig:conformer}
\end{figure}

\subsection{Experimental setup}

Both baseline and proposed models were trained for a maximum of 20 epochs with a batch size of 128, employing an early stopper (patience level of 3) based on development set MSE metric.

All experiments were conducted on a server equipped with an Intel Core i7-4770 CPU (3.40 GHz, 8 cores, 16 threads) and an NVIDIA TITAN Xp GPU (12 GB VRAM) with 32GB RAM.  The system ran Ubuntu 18.04.6 LTS (kernel 5.15.0-41-generic). Deep learning computations were performed using TensorFlow 2.18.0 with Keras 3.8.0, CUDA 11.4, and NVIDIA driver 470.129.06.

\subsection{Baseline}
As a baseline, we trained a standard 2D-CNN architecture for each speaker to map UTIF to an 80-dimensional mel spectrogram using the open-access implementation\footnote{\href{https://github.com/BME-SmartLab/UTI-to-STFT}{https://github.com/BME-SmartLab/UTI-to-STFT}} provided by Csapó et al.~\cite{csapo20b_interspeech}. The speaker-specific model consisted of 4.09 million trainable parameters.

\subsection{Conformer Base}
We implemented the exact same sandwich-style structure of the original conformer architecture (four modules followed by layer normalization). We arranged the full 'Conformer Base' network with linear layers around the core conformer block (Figure~\ref{fig:conformer}). To ensure compatibility with the dimensions of the target mel spectrogram, the output of the final linear layer was flattened before being passed to the output layer.


The input of our network is UTIF as a sequential data, where each of the frames with 64 beam lines represents 12ms of the recording.  We processed the input by treating each of these beam lines as a separate time step, resulting in a sequence length of 64 × (time dimension of the recording), as illustrated in Figure~\ref{fig:pipeline}. Utilizing the ability of the conformer block for generalizing the global and local dependencies, we fed the network with a 64×128 partition of the sequence starting from the beginning till the ending in a consecutive order. This maintained frame-level learning, where each frame corresponds to one time step of the 80-dimensional mel spectrogram output.


The conformer block was configured with an encoder dimension of 256, 32 attention heads, a convolution kernel size of 31, and a feedforward expansion factor of 3 while keeping the rest of the hyperparameters as in the original structure. The model was compiled using the AdamW optimizer with a custom cosine decay restarts learning rate scheduler. The initial learning rate was 0.0001, the first cycle lasted 100 steps, and subsequent cycles increased in length by a factor of 5. The maximum learning rate decayed by a factor of 0.9 each cycle, with a minimum learning rate of 0.00001. This proposed acoustic feature generator had 2.66 million trainable parameters. The training time per speaker was approximately 30\% of that required by the baseline model.


\subsection{Conformer with bi-LSTM}
Earlier research on speech synthesis from three-axis accelerometer signal by Kwon et al.~\cite{KWON2024109090}, has utilized conformer-based network towards SSI. Their network consisted of four conformer blocks followed by two layers of bi-LSTM. As bi-LSTM can process sequence data more effectively by simultaneously considering both past and future information, its application on time-series data such as an accelerometer signal showed efficient performance. 



Inspired by this implementation, we also incorporated bi-LSTM layers into our 'Conformer Base' model, creating the 'Conformer with bi-LSTM' model. We augmented the conformer block with two bi-LSTM layers following layer normalization (Figure~\ref{fig:conformer}). The rest of the network architecture remained the same as in the 'Conformer Base' layout. With 5.35 million trainable parameters, the model was trained in approximately 80\% of the time required by the baseline model per speaker.

\begin{table}[t]
\caption{Mean Squared Error results on the test set for each speaker.} \label{tab:tab_mse}
\centering
\renewcommand{\arraystretch}{1.1} 
\resizebox{\linewidth}{!}{
\begin{tabular}{|l||c|c|c|c|}
\hline
{\bf Neural network} & \multicolumn{4}{c|}{\bf Speaker ID} \\
\cline{2-5}
{\bf models} & 01fi & 02fe & 03mn & 04me  \\
\hline\hline
Baseline   & 0.464 & 0.623 & 0.395 & 0.484 \\
\hline
Conformer  & 0.511  & 0.618  & 0.462  & 0.524  \\
Base                & (p = 0.162) & (p = 0.521) & (p = 0.026) & (p = 0.121) \\
\hline
Conformer   & 0.482 & 0.581 & 0.378 & 0.449 \\
with bi-LSTM                & (p = 0.678) & (p = 1.000) & (p = 0.678) & (p = 0.141) \\
\hline 
\end{tabular}
}
\end{table}

\subsection{Speech synthesis}
For the speech synthesis component of the proposed ultrasound-to-speech conversion system depicted in Figure~\ref{fig:pipeline}, we employed the HiFi-GAN neural vocoder~\cite{kong2020hifi}. We chose the first variation of the openly accessible pre-trained model on the multi-speaker VCTK dataset (folder name: VCTK\_V1)\footnote{\href{https://github.com/jik876/hifi-gan}{https://github.com/jik876/hifi-gan}}, due to its higher synthesized audio quality generation ability.

\section{Results}

\subsection{Preliminary consideration}

Given the speaker-specific nature of the models and the variability in UTIF quality across subjects~\cite{toth23_interspeech}, a speaker-wise analysis of UTS systems is necessary. With four speakers, we examined objective measurements individually to identify speaker-wise trends. However, subjective evaluations were conducted both per speaker and averaged across all speakers to provide both specific and general performance insights.

\subsection{Mean Squared Error (MSE)}

To measure the performance of the proposed networks in comparison to the baseline, we calculated MSE values between the generated and the original mel spectrograms from the test set. We measured the statistical significance of differences between MSE scores of proposed and baseline methods within one speaker. We used the Mann-Whitney U test~\cite{mann1947onatest} to compare sentence-level MSE scores between models for each speaker with a 95\% confidence level. The mean MSE results of ten utterances per speaker are shown in Table~\ref{tab:tab_mse} with their respective significance scores (p-values); a p-value $\ge$ 0.05 suggests that there is not enough evidence to say that there is a significant difference. 

Looking at the results, we can clearly say that the performance of the proposed and the baseline networks are similar to each other for all the speakers. However, for speaker "03mn", the mean MSE value of the `Conformer Base' model is noticeably higher than the baseline, reaching the level of statistical significance ($p = 0.026$).


\subsection{Mel-Cepstral Distortion (MCD)}

\begin{table}[t]
\caption{Mel-Cepstral Distortion values (dB) on the test set per speaker.} \label{tab:tab_mcd}
\centering
\renewcommand{\arraystretch}{1.1} 
\resizebox{\linewidth}{!}{
\begin{tabular}{|l||c|c|c|c|}
\hline
{\bf Neural network} & \multicolumn{4}{c|}{\bf Speaker ID} \\
\cline{2-5}
{\bf models} & 01fi & 02fe & 03mn & 04me  \\
\hline\hline
Baseline   & 3.221 & 3.009 & 3.641 & 3.172  \\
\hline
Conformer   & 3.517 & 3.121 & 4.133 & 3.465 \\
Base            & (p = 0.001) & (p = 0.104) & (p = 0.001) & (p = 0.001) \\
\hline
Conformer   & 3.253 & 3.037 & 3.704 &  3.258 \\
with bi-LSTM            & (p = 0.241) & (p = 0.791) & (p = 0.186) & (p = 0.002) \\
\hline
\end{tabular}
}
\end{table}

We also used the MCD metric to objectively assess the quality of the synthesized speech from our proposed and baseline UTS systems for each speaker, which summarizes the spectral differences between two waveforms. MCD scores were obtained by utilizing a publicly available implementation\footnote{\href{https://github.com/ttslr/python-MCD}{https://github.com/ttslr/python-MCD}}, with lower values indicating better synthesis quality. 

Table~\ref{tab:tab_mcd} shows mean MCD measurement results in dB with the corresponding significance scores. We can see that the performance of the proposed UTS systems varies across speakers according to the MCD values. The conformer-based networks both performed on the level of the baseline for speaker "02fe", but exhibited significantly higher MCD values for speaker "04me". While the performance of 'Conformer with bi-LSTM' for the remaining speakers was also similar to the baseline 2D-CNN, 'Conformer Base' performed significantly lower performance with higher MCD values.

\begin{figure}[t]
  \centering
  \includegraphics[width=\linewidth]{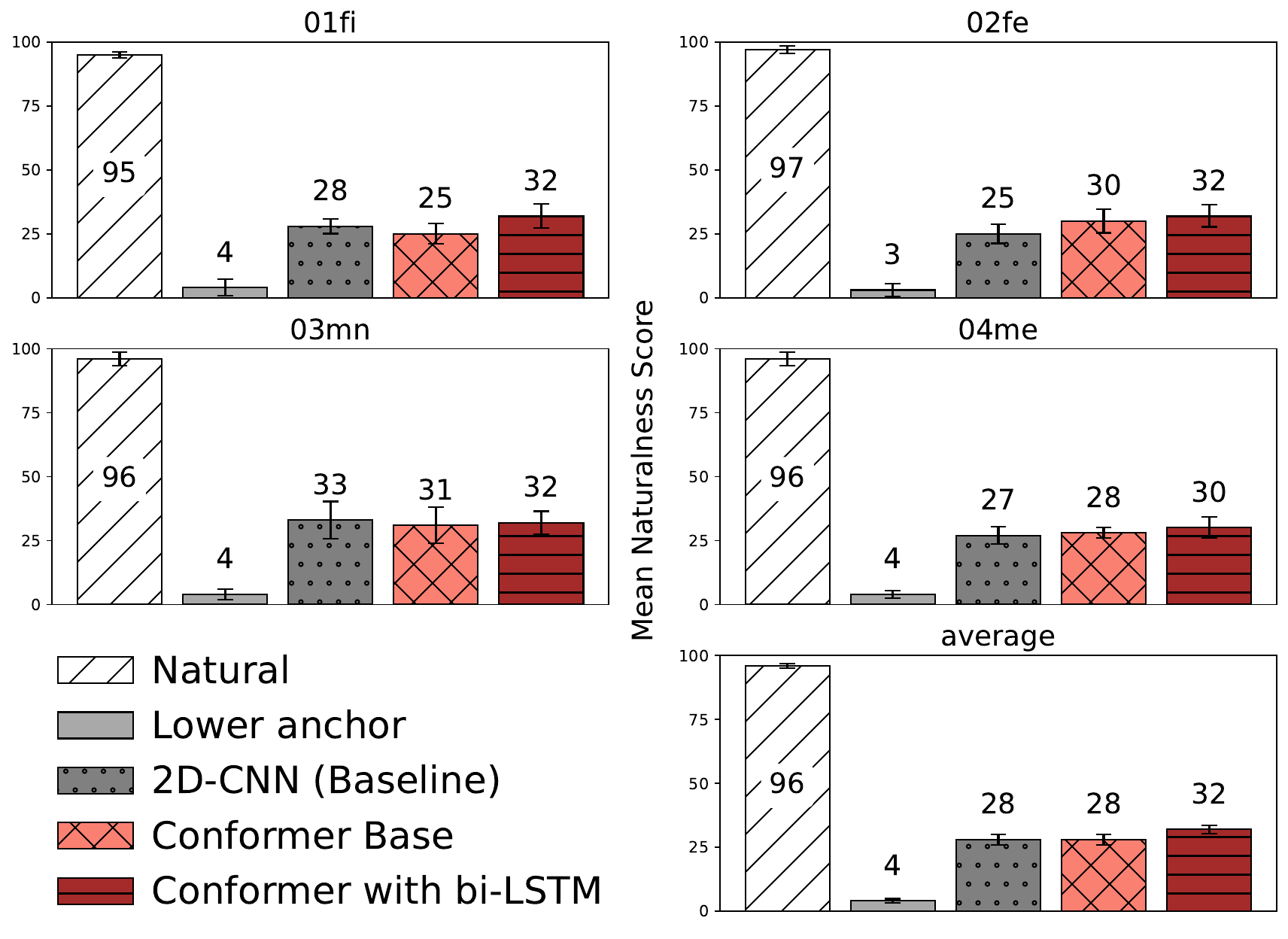}
  \caption{Results of the MUSHRA listening test with respect to naturalness, speaker by speaker (first two rows) and on average (bottom-right corner). The error bars show the 95\% confidence intervals.}
  \label{fig:mushra}
\end{figure}

\subsection{Subjective evaluation}

To assess the perceived naturalness of the synthesized speech, we also conducted a MUSHRA listening test~\cite{ITU-R-BS.1534}. Listeners rated the naturalness of each synthesized utterance relative to the corresponding natural speech reference, using a scale from 0 (very unnatural) to 100 (very natural). Five utterances were randomly selected from the test set per each speaker (20 utterances in total). Our aim was to compare the natural sentences with the synthesized ones of the baseline, the proposed approaches and a lower anchor, which is simply a version of the original speech sample with added white noise at a 0.0005 level. The order of presentation was randomized for each of the 27 non-native English listeners (14 female, 13 male; aged 20–55). On average, the test took 18 minutes to complete. 

Figure~\ref{fig:mushra} displays the mean naturalness ratings for each system, showing both speaker-specific results and the overall average. The lower anchor version achieved the lowest scores, while the natural sentences were rated the highest, as expected. Clearly, the findings presented in the table indicate that, at least one of the proposed methods offer a potential increase in the naturalness of synthesized speech over the baseline for three out of four speakers, the only exception being speaker "03mn". 

Based on the average naturalness values, there is a clear preference for the proposed UTS system utilizing the `Conformer with biLSTM' model. The difference in performance between the `Conformer with biLSTM' and the baseline model for all the utterances across subjects is statistically significant, with a $p$ value of 0.037. In contrast, the `Conformer Base' model showed no significant difference compared to the baseline, with a $p$ value of 0.698 on average across speakers, indicating a high level of similarity. 

In addition to comparing the proposed methods to the baseline, we also assessed the performance difference between the two proposed models. A significant difference in naturalness (p = 0.015) was observed between the two proposed systems, with the UTS system using `Conformer with bi-LSTM' performing better than the `Conformer Base' UTS systems.

\section{Conclusions and discussion}

\begin{figure}[t]
  \centering
  \includegraphics[width=0.97\linewidth]{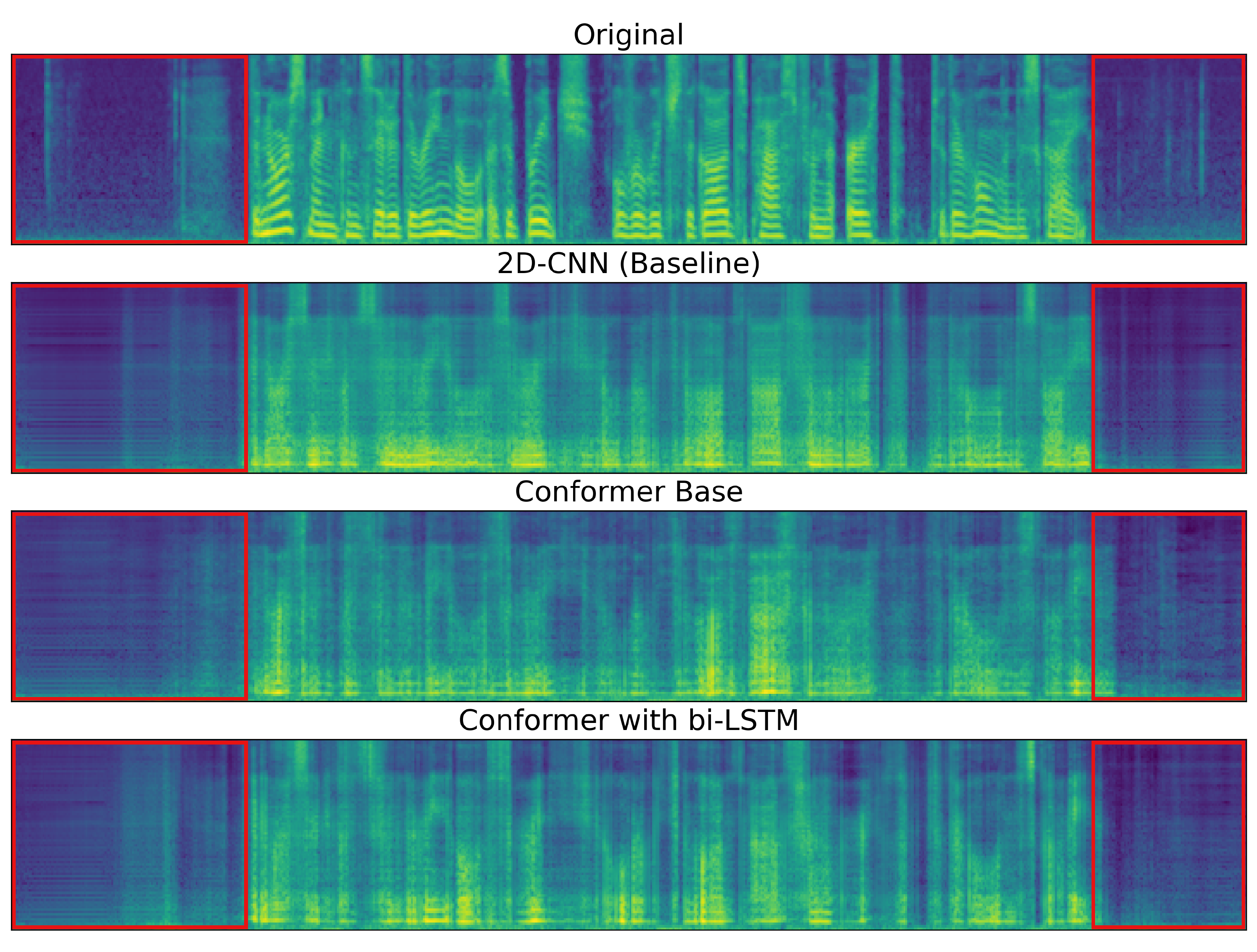}
  \caption{Mel spectrogram representation of synthesized samples using proposed and baseline ultrasound-to-speech system, in comparison to original form ("014\_xaud" utterance by speaker "01fi").}
  \label{fig:mel}
\end{figure}

In the ultrasound-to-speech 
area there is a need for more effective acoustic feature generation techniques. In this work we presented two models based on the Conformer architecture for 
mapping 
ultrasound tongue images to 80-dimensional mel spectrograms, 
and 
used the VCTK\_V1 variant of the HiFi-GAN vocoder for the speech synthesis step. The two conformer-based networks were compared to a 2D-CNN baseline.

The performance of the ultrasound-to-mel mapping step was evaluated using the MSE metric. The results displayed a general tendency that the effectiveness of the proposed methods was similar to the baseline. However, the mean MCD values between the generated and the original speech signals did not follow the same trend, but showed a speaker-dependent aspect of our UTS system. Finally, a MUSHRA listening test indicated a strong preference for the `Conformer with bi-LSTM' model, while `Conformer Base' performed similarly as the baseline. Importantly, both proposed models had fewer parameters and required less training time than the baseline 2D-CNN, promising a better potential for real-time applications.

We visually compared the original and all generated mel spectrograms for utterance "014\_xaud" (speaker "01fi," from the 
test set) to understand the conflicting MSE and MCD results (see Figure~\ref{fig:mel}). 
We can say that the proposed methods (especially `Conformer with biLSTM') capture more spectral details (e.g. formant structure) than the baseline models. However, this visual improvement in 
the mel spectrograms
did not lead to improved objective scores. 
Our hypothesis is that the ability of the proposed models to generate the silent parts (indicated by red boxes) is not more accurate than the baseline.
This might be the reason in the discrepancy of objective results, which 
we plan to analyze further as a future work.

Considering both subjective evaluation results and the computational advantages, we conclude that both proposed conformer-based models offer viable alternatives for ultrasound-to-speech conversion, with a clear preference for the `Conformer with bi-LSTM' system.

For reproducibility, the full code and synthesized speech samples are available at https://doi.org/10.5281/zenodo.15544868.

\section{Acknowledgements}
\ifinterspeechfinal
This study was supported by the NRDI Office of the Hungarian Ministry of Innovation and Technology (grant TKP2021-NVA-09), and within the framework of the Artificial Intelligence National Laboratory Program (RRF-2.3.1-21-2022-00004) and the European Union’s HORIZON Research and Innovation Programme under grant agreement No 101120657, project ENFIELD (European Lighthouse to Manifest Trustworthy and Green AI).

This work is dedicated to the memory of Dr. Tamás Gábor Csapó, whose inspiration and influence continue to guide us. May he rest in peace.

\else
Placeholder placeholder placeholder placeholder placeholder placeholder placeholder placeholder placeholder placeholder placeholder placeholder placeholder placeholder placeholder placeholder placeholder placeholder placeholder placeholder placeholder placeholder placeholder placeholder placeholder placeholder placeholder placeholder placeholder placeholder 
\fi

\bibliographystyle{IEEEtran}
\bibliography{mybib}

\begin{thebibliography}{10}
\providecommand{\url}[1]{#1}
\csname url@samestyle\endcsname
\providecommand{\newblock}{\relax}
\providecommand{\bibinfo}[2]{#2}
\providecommand{\BIBentrySTDinterwordspacing}{\spaceskip=0pt\relax}
\providecommand{\BIBentryALTinterwordstretchfactor}{4}
\providecommand{\BIBentryALTinterwordspacing}{\spaceskip=\fontdimen2\font plus
\BIBentryALTinterwordstretchfactor\fontdimen3\font minus \fontdimen4\font\relax}
\providecommand{\BIBforeignlanguage}[2]{{%
\expandafter\ifx\csname l@#1\endcsname\relax
\typeout{** WARNING: IEEEtran.bst: No hyphenation pattern has been}%
\typeout{** loaded for the language `#1'. Using the pattern for}%
\typeout{** the default language instead.}%
\else
\language=\csname l@#1\endcsname
\fi
#2}}
\providecommand{\BIBdecl}{\relax}
\BIBdecl

\bibitem{zhu2021towards}
M.~Zhu, H.~Zhang, X.~Wang, X.~Wang, Z.~Yang, C.~Wang, O.~W. Samuel, S.~Chen, and G.~Li, ``Towards optimizing electrode configurations for silent speech recognition based on high-density surface electromyography,'' \emph{Journal of Neural Engineering}, vol.~18, no.~1, p. 016005, 2021.

\bibitem{sEMG_2}
S.~Khan, K.~U. Ali, A.~R.~A. Qadri, and M.~Rizvi, ``Silent speech interfaces: Non-invasive neuromuscular signal processing for assistive communication,'' in \emph{Proceedings of ICAIC}, 2025, pp. 1--6.

\bibitem{trencsenyi2024ultrasound}
R.~Trencs\'enyi and L.~Czap, ``Ultrasound-and {MRI}-based speech synthesis applying neural networks,'' in \emph{Proceedings of ICCC}.\hskip 1em plus 0.5em minus 0.4em\relax IEEE, 2024, pp. 1--6.

\bibitem{toutios2016articulatory}
A.~Toutios, T.~Sorensen, K.~Somandepalli, R.~Alexander, and S.~S. Narayanan, ``Articulatory synthesis based on real-time magnetic resonance imaging data,'' in \emph{Proceedings of Interspeech}, 2016, pp. 1492--1496.

\bibitem{ribeiro21_interspeech}
M.~S. Ribeiro, A.~Eshky, K.~Richmond, and S.~Renals, ``Silent versus modal multi-speaker speech recognition from ultrasound and video,'' in \emph{Proceedings of Interspeech}, 2021, pp. 641--645.

\bibitem{pandey2024melder}
L.~Pandey and A.~S. Arif, ``Melder: The design and evaluation of a real-time silent speech recognizer for mobile devices,'' in \emph{Proceedings of the CHI Conference on Human Factors in Computing Systems}, 2024, pp. 1--23.

\bibitem{hueber2016statisticalconversion}
T.~Hueber and G.~Bailly, ``Statistical conversion of silent articulation into audible speech using full-covariance {HMM},'' \emph{Computer, Speech \& Language}, vol.~36, pp. 274--293, 2016.

\bibitem{guo2024multi}
M.~Guo, J.~Wei, R.~Zhang, Y.~Zhao, and Q.~Fang, ``Multi-modal co-learning for silent speech recognition based on ultrasound tongue images,'' \emph{Speech Communication}, vol. 165, p. 103140, 2024.

\bibitem{csapo17_interspeech}
T.~G. Csapó, T.~Grósz, G.~Gosztolya, L.~Tóth, and A.~Markó, ``{DNN}-based ultrasound-to-speech conversion for a silent speech interface,'' in \emph{Proceedings of Interspeech}, 2017, pp. 3672--3676.

\bibitem{article_csapo_lstm}
E.~Moliner and T.~Csapó, ``Ultrasound-based silent speech interface using convolutional and recurrent neural networks,'' \emph{Acta Acustica united with Acustica}, vol. 105, 07 2019.

\bibitem{saha2020ultra2speech}
P.~Saha, Y.~Liu, B.~Gick, and S.~Fels, ``Ultra2speech -- {A} deep learning framework for formant frequency estimation and tracking from ultrasound tongue images,'' in \emph{Proceedings of MICCAI}, Lima, Peru, Oct 2020, pp. 473--482.

\bibitem{kimura2019sottovoce}
N.~Kimura, M.~Kono, and J.~Rekimoto, ``Sotto{V}oce: {A}n ultrasound imaging-based silent speech interaction using deep neural networks,'' in \emph{Proceedings of the 2019 CHI Conference on Human Factors in Computing Systems}, 2019, pp. 1--11.

\bibitem{csapo20b_interspeech}
T.~G. Csapó, C.~Zainkó, L.~Tóth, G.~Gosztolya, and A.~Markó, ``Ultrasound-based articulatory-to-acoustic mapping with {WaveGlow} speech synthesis,'' in \emph{Proceedings of Interspeech}, 2020, pp. 2727--2731.

\bibitem{NIPS2017_3f5ee243}
A.~Vaswani, N.~Shazeer, N.~Parmar, J.~Uszkoreit, L.~Jones, A.~N. Gomez, L.~u. Kaiser, and I.~Polosukhin, ``Attention is all you need,'' in \emph{Proceedings of NIPS}, vol.~30.\hskip 1em plus 0.5em minus 0.4em\relax Curran Associates, Inc., 2017.

\bibitem{dosovitskiy2020image}
A.~Dosovitskiy, ``An image is worth 16x16 words: {T}ransformers for image recognition at scale,'' in \emph{Proceedings of NIPS}, 2020.

\bibitem{genomics_transformer}
J.~Clauwaert and W.~Waegeman, ``Novel transformer networks for improved sequence labeling in genomics,'' \emph{IEEE/ACM Transactions on Computational Biology and Bioinformatics}, vol.~19, no.~1, pp. 97--106, 2022.

\bibitem{moritz2020streaming}
N.~Moritz, T.~Hori, and J.~Le, ``Streaming automatic speech recognition with the transformer model,'' in \emph{Proceedings of ICASSP}.\hskip 1em plus 0.5em minus 0.4em\relax IEEE, 2020, pp. 6074--6078.

\bibitem{SONG2023104298}
R.~Song, X.~Zhang, X.~Chen, X.~Chen, X.~Chen, S.~Yang, and E.~Yin, ``Decoding silent speech from high-density surface electromyographic data using transformer,'' \emph{Biomedical Signal Processing and Control}, vol.~80, p. 104298, 2023.

\bibitem{gulati20_interspeech}
A.~Gulati, J.~Qin, C.-C. Chiu, N.~Parmar, Y.~Zhang, J.~Yu, W.~Han, S.~Wang, Z.~Zhang, Y.~Wu, and R.~Pang, ``Conformer: Convolution-augmented transformer for speech recognition,'' in \emph{Proceedings of Interspeech}, 2020, pp. 5036--5040.

\bibitem{KWON2024109090}
J.~Kwon, J.~Hwang, J.~E. Sung, and C.-H. Im, ``Speech synthesis from three-axis accelerometer signals using conformer-based deep neural network,'' \emph{Computers in Biology and Medicine}, vol. 182, p. 109090, 2024.

\bibitem{ribeiro2021talasynchronized}
M.~S. Ribeiro, J.~Sanger, J.-X.~X. Zhang, A.~Eshky, A.~Wrench, K.~Richmond, and S.~Renals, ``{TaL}: {A} synchronised multi-speaker corpus of ultrasound tongue imaging, audio, and lip videos,'' in \emph{Proceedings of SLT}, Shenzhen, China, 2021, pp. 1109--1116.

\bibitem{kong2020hifi}
J.~Kong, J.~Kim, and J.~Bae, ``Hi{F}i-{GAN}: {G}enerative adversarial networks for efficient and high fidelity speech synthesis,'' \emph{Proceedings of NIPS}, vol.~33, pp. 17\,022--17\,033, 2020.

\bibitem{toth23_interspeech}
L.~Tóth, A.~{Honarmandi Shandiz}, G.~Gosztolya, and T.~G. Csapó, ``Adaptation of tongue ultrasound-based silent speech interfaces using {S}patial {T}ransformer {N}etworks,'' in \emph{Proceedings of Interspeech}, 2023, pp. 1169--1173.

\bibitem{mann1947onatest}
H.~B. Mann and D.~R. Whitney, ``On a test of whether one of two random variables is stochastically larger than the other,'' \emph{Annals of Mathematical Statistics}, vol.~18, no.~1, pp. 50--60, 1947.

\bibitem{ITU-R-BS.1534}
{International Telecommunication Union}, ``{ITU-R Recommendation BS.1534: Method for the Subjective Assessment of Intermediate Audio Quality},'' Tech. Rep. BS.1534, 2001.

\end{thebibliography}

\end{document}